# PREDICTIVE COMPARATIVE QSAR ANALYSIS OF SULFATHIAZOLE ANALOGUES AS MYCOBACTERIUM TUBERCULOSIS H37RV INHABITORS

**Doreswamy and *Chanabasyya M. Vastrad**

Department of Computer Science Mangalore University, Mangalagangotri-574 199, Karnataka, INDIA
Email: doreswamyh@yahoo.com , *channu.vastrad@gmail.com  Tel: +91-9480073398

**ABSTRACT:**
Antitubercular activity of Sulfathiazole Derivitives series were subjected to Quantitative Structure Activity Relationship (QSAR) Analysis with an attempt to derive and understand a correlation between the Biologically Activity as dependent variable and various descriptors as independent variables. QSAR models generated using 28 compounds. Several statistical regression expressions were obtained using Partial Least Squares (PLS) Regression ,Multiple Linear Regression (MLR) and Principal Component Regression (PCR) methods. The among these methods, Partial Least Square Regression (PLS) method has shown very promising result as compare to other two methods. A QSAR model was generated by a training set of 18 molecules with correlation coefficient r ( $r^2$ ) of 0.9191 , significant cross validated correlation coefficient ($q^2$) of 0.8300 , F test of 53.5783 , $r^2$ for external test set ($pred\_r^2$) -3.6132, coefficient of correlation of predicted data set ($pred\_r^2se$) 1.4859 and degree of freedom 14 by Partial Least Squares Regression Method.

## [I] INTRODUCTION

Tuberculosis in humans is mainly caused by Mycobacterium tuberculosis.The infection is transmitted by respirable droplets generated during forceful expiratory manoeuvres such as coughing. Tuberculosis infection can be either active or latent[1] . The World Health Organization (WHO) estimates that within the next 20 years about 30 million people will be infected with the bacillus [2-3]. The clinical management of TB has relied heavily on a limited number of drugs such as Isonicotinic acid, Hydrazide, Rifampicin, Ethambutal, Streptomycin, Ethionamide, Pyrazinamide, Fluroquinolones etc [4]. However with the advent of these chemotherapeutic agents the spread of TB has not been eradicated completely because of prolonged treatment schedules There is now recognition that new drugs to treat TB are urgently required, specifically for use in shorter treatment regimens than are possible with the current agents and which can be employed to treat multi-drug resistant and latent disease[5].

Sulfathiazoles exhibit potent in vitro and in vivo antimycobacterial activity [6]. There is also a considerable effort to discover and develop newer sulfathiazoles, and some of them might have value in the treatment of TB [7]. Cheminformatics[26] and computer-aided drug design (CADD) are expected to contribute to a possible solution for the perilous situation regarding this infectious disease by assisting in the rapid identification of novel effective anti-TB agents. An alternative way for overcoming the absence of experimental measurements for biological systems is based on the activity to formulate quantitative structure activity relationship (QSAR) [8] . QSAR models are mathematical equations constructing a relationship between chemical structures and biological activities. These models have another ability, which is providing a deeper knowledge about the mechanism of biological activity. In the first step of a typical QSAR study one needs to find a set of molecular descriptors with the higher impact on the biological activity of interest [9]. A wide range of descriptors[10] has been used in QSAR modeling. These descriptors[11] have been classified into different categories, including constitutional, geometrical, topological, quantum chemical and so on. Using such an approach one could predict the activities of newly designed compounds before a decision is being made whether these compounds should be really synthesized and tested. In this work, we attempt to compare the performance of Partial Least Squares(PLS) based QSAR models with the results produced by Multi Linear Regression(MLR ) and Principal Component Regression (PCR) techniques to find structural requirements for further improved antitubercular activity.

## [II] MATERIALS AND METHODS
### 2.1 Molecular Data Sets

A series of 28 molecules belonging to derivatives for Mycobacterium tuberculosis(H37Rv) inhibitors were taken from large Antituberculosis drug discovry databases[12] using Substructure mining tool Schrodinger Canvas 2010(Trial version)[13]. All molecules were processed by the Vlife MDS [14] - 2D coordinates of atoms were recalculated counter ions and salts were removed from molecular structures, molecules were neutralized, mesomerized and aromatized. Data sets were then filtered from duplicates. The 2D-QSAR models were generated using a training set of 18 molecules. Predictive power of the resulting models was evaluated by a test set of 10 molecules with uniformly distributed biological activities. The observed selection of test set molecules was made by considering the fact that test set molecules represents a range of biological activity similar to the training set. The observed and predicted biological activities of the training and test set molecules are presented in Table 1.

### 2.2 Biological Activity Data

For the development of QSAR models of Sulfathiazoles , in vitro antitubercusis activity in terms of half maximal inhibitory concentration IC50 (μM) against (H37Rv) strains were taken from the Antituberculosis drug discovry databases[12]. The IC50 summary data contains only molecules that have at least exhibited some activity. The biological activity data (IC50) were converted in to pIC50 according to the formula pIC50 = (-log (IC50 X $10^{-6}$)) was used as dependent variable, thus correlating the data linear to the free energy change.

### 2.3 Descriptor calculation

The VLife MDS tool was employed for the calculation of different descriptors including topological index (J), connectivity index (x), radius of gyration (RG), moment of inertia, Wiener index(W), balaban index(J), centric index , hosoya index (Z), information based indices, XlogP, logP , hydrophobicity, elemental count, path count, chain count, pathcluster count, molecular connectivity index (chi), kappa values, electro topological state indices, electrostatic surface properties, dipole moment, polar surface area(PSA), alignment independent descriptor (AI)[11,14] . The calculated descriptors were gathered in a data matrix. The preprocessing of the independent variables (i.e., descriptors) was done by removing invariable (constant column) and cross-correlated descriptors (with r = 0.99). which resulted in total 156, 125 and 162 descriptors for MLR, PCR and PLS respectively to be used for QSAR analysis.

**2.4 Selection of Training and Test Set**

The dataset of 28 molecules was divided into training and test set by Sphere Exclusion (SE)[15-16] method. In this method initially data set were divided into training and test set using sphere exclusion method. In this method dissimilarity value provides an idea to handle training and test set size. It needs to be adjusted by trial and error until a desired division of training and test set is achieved. Increase in dissimilarity value results in increase in number of molecules in the test set. This method is used for MLR, PCR and PLS model with pIC50 activity field as dependent variable and various 2D descriptors calculated for the molecules as independent variables.

**2.5 Model Validation**

Validation [17-18] is a crucial aspect of quantitative structure–activity relationship (QSAR) modeling. This is done to test the internal stability and predictive ability of the QSAR models. Developed QSAR models were validated by the following procedure.

**2.5.1 Internal Validation**

Internal validation was carried out using leave-one-out ( LOO-$Q^2$) method. For calculating $q^2$, each molecule in the training set was eliminated once and the activity of the eliminated molecule was predicted by using the model developed by the remaining molecules. The $q^2$ was calculated using the equation which describes the internal stability of a model.

$$Q^2 = 1 - \frac{\Sigma(Y_{pred} - Y_{obs})^2}{\Sigma(Y_{obs} - Y_{mean})^2} \quad \text{------ (1)}$$

In Eq. (1), $Y_{pred}$ and $Y_{obs}$ indicate predicted and observed activity values respectively and $Y_{mean}$ indicate mean activity value. A model is considered acceptable when the value of $Q^2$ exceeds 0.5.

**2.5.2 External Validation**

For external validation, the activity of each molecule in the test set was predicted using the model developed by the training set. The $pred\_r^2$ value is calculated as follows.

$$pred\_r^2 = \frac{\Sigma(Y_{pred(Test)} - Y_{Test})^2}{\Sigma(Y_{Train} - Y_{mean(Train)})^2} \quad \text{------ (2)}$$

In Eq (2) $Y_{pred(Test)}$ and $Y_{Test}$ indicate predicted and observed activity values respectively of the test set compounds and $Y_{Train}$ indicates mean activity value of the training set. For a predictive QSAR model, the value of $pred\_r^2$ should be more than 0.5.

**2.5.3 Randomization Test**

Randomization test or Y-scrambling is important popular mean of statistical validation. To evaluate the statistical significance of the QSAR model for an actual dataset, one tail hypothesis testing was used. The robustness of the models for training sets was examined by comparing these models to those derived for random datasets. Random sets were generated by rearranging the activities of the molecules in the training set. The statistical model was derived using various randomly rearranged activities (random sets) with the selected descriptors and the corresponding $q^2$ were calculated. The significance of the models hence obtained was derived based on a calculated $Z_{score}$.

A Z score value is calculated by the following formula:

$$Z_{score} = \frac{(h - \mu)}{\sigma} \quad \text{---------------- (3)}$$

Where h is the $q^2$ value calculated for the actual dataset, μ the average $q^2$, and s is its standard deviation calculated for various iterations using models build by different random datasets. The probability (a) of significance of randomization test is derived by comparing $Z_{score}$ value with $Z_{score}$ critical value as reported, if $Z_{score}$ value is less than 4.0; otherwise it is calculated by the formula as given in the literature. For example, a $Z_{score}$ value greater than 3.10 indicates that there is a probability (a) of less than 0.001 that the QSAR model constructed for the real dataset is random. The randomization test suggests that all the developed models have a probability of less than 1% that the model is generated by chance.

### 2.6 Multiple Linear Regression (MLR) Analysis

MLR is a method used for modeling linear relationship between a dependent variable Y (pIC50) and independent variable X (2D descriptors). MLR is based on least squares: the model is fit such that sum-of-squares of differences of observed and a predicted value is minimized. MLR estimates values of regression coefficients ($r^2$) by applying least squares curve fitting method. The model creates a relationship in the form of a straight line (linear) that best approximates all the individual data points. In regression analysis, conditional mean of dependant variable (pIC50) Y depends on (descriptors) X. MLR analysis extends this idea to include more than one independent variable. Regression equation takes the form.

$$Y = b_1x_1 + b_2x_2 + b_3x_3 \text{ --------- (4)}$$

where Y is dependent variable, 'b's are regression coefficients for corresponding 'x's (independent variable), 'c' is a regression constant or intercept [19,25].

### 2.7 Principal Component Regression (PCR) Analysis

PCR is a data compression method based on the correlation among dependent and independent variables. PCR provides a method for finding structure in datasets. Its aim is to group correlated variables, replacing the original descriptors by new set called principal components (PCs). These PCs uncorrelated and are built as a simple linear combination of original variables. It rotates the data into a new set of axes such that first few axes reflect most of the variations within the data. First PC (PC1) is defined in the direction of maximum variance of the whole dataset. Second PC (PC2) is the direction that describes the maximum variance in orthogonal subspace to PC1. Subsequent components are taken orthogonal to those previously chosen and describe maximum of remaining variance, by plotting the data on new set of axes, it can spot major underlying structures automatically. Value of each point, when rotated to a given axis, is called the PC value. PCA selects a new set of axes for the data. These are selected in decreasing order of variance within the data. Purpose of principal component PCR is the estimation of values of a dependent variable on the basis of selected PCs of independent variables [21].

### 2.8 Partial Least Squares (PLS) Regression Analysis

PLS analysis is a popular regression technique which can be used to relate one or more dependent variable (Y) to several independent (X) variables. PLS relates a matrix Y of dependent variables to a matrix X of molecular structure descriptors. PLS is useful in situations where the number of independent variables exceeds the number of observation, when X data contain colinearties or when N is less than 5 M, where N is number of compound and M is number of dependant variable. PLS creates orthogonal components using existing correlations between independent variables and corresponding outputs while also keeping most of the variance of independent variables. Main aim of PLS regression is to predict the activity (Y) from X and to describe their common structure [22,23]. PLS is probably the least restrictive of various multivariate extensions of MLR model. PLS is a method for constructing predictive models when factors are many and highly collinear

### 2.9 Evaluation of the QSAR Models

The Developed QSAR models are evaluated using the following statistical measures: n (Number of compounds in regression); K (Number of variables(desriptors)); DF (Degree of freedom); optimum component ( number of optimums); $r^2$ ( the squared correlation coefficient); F test (Fischer's Value) for statistical significance; $q^2$ (cross-validated correlation coefficient); $pred\_r^2$ ($r^2$ for external test set); $Z_{score}$ ( Z score calculated by the randomization test); $Best\_ran\_r^2$ (highest $r^2$ value in the randomization test) ; $Best\_ran\_q^2$ ( highest $q^2$ value in the randomization test) ; α ( statistical significance parameter obtained by the randomization test). The regression coefficient $r^2$ is a relative measure of fit by the regression equation. It represents the part of the variation in the observed data is explained by the regression. However , a QSAR model is considered to be predictive , if the following conditions are satisfied: $r^2 > 0.6$ , $q^2 > 0.6$ and $pred\_r^2 > 0.5$ [24] . The F-test refects the ratio of variance explained by the model and variance due to the error in the regression. High values of the F-test indicate that model is statisticaly significant. The low standard error of $pred\_r^2se$, $q^2\_se$ and $r^2\_se$ shows absoute quility of the fittness of the model. The cross-correlation limit was set at 0.5.

**[III] RESULTS**
Training set of 18 and 10 of test set of Sulfathiazoles having different substitution were employed.
**3.1 Generation of QSAR Models**
**3.1.1 Model – 1 Partial Least Squares (PLS) Regression Analysis**
The compounds were subjected to under goes PLS method to developed QSAR models by using Simulated anealining variable selection mode. Model - 1 is having following QSAR equation 5 with 5 variables.
pIC50 = -0.0317(PolarSurfaceAreaExcludingPandS) - 0.00001(MomInertiaX) - 0.5204(slogp)
 - 0.6920(SaaScount)-0.0562(SsOHE-index) + 11.5501 ---- (5)

The model -2 gave correlation coefficient ( $r^2$) of 0.9199, significant cross validated correlation coefficient ( $q^2$) of 0.8300 , F test of 53.5783 and degree of freedom 14. The model is validated by $\alpha\_ran\_r^2$= 0.00000, $\alpha\_ran\_q^2$ = 0.00000, $best\_ran\_r^2 = 0.48024$, $best\_ran\_q^2$= -0.07412, $Z\_score\_ran\_r^2$ = 5.55165 and $Z\_score\_ran\_q^2$ = 5.41451. The randomization test suggests that the developed model have a probability of less than 1% that the model is generated by chance. Statistical data is shown in Table 2. The plot of observed vs. predicted activity is shown in Figure 1. The descriptors which contribute for the pharmacological action are shown in Figure 2.

| Parameters | PLS | MLR | PCR |
|---|---|---|---|
| N | 28 | 28 | 28 |
| DF | 14 | 13 | 14 |
| $r^2$ | 0.9199 | 0.8647 | 0.8088 |
| $q^2$ | 0.8300 | 0.7692 | 0.6715 |
| F-test | 53.5783 | 20.7628 | 19.7379 |
| $best\_ran\_r^2$ | 0.48024 | 0.44613 | 0.25466 |
| $best\_ran\_q^2$ | 0.07412 | -0.02584 | 0.13938 |
| $Z\_score\_ran\_r^2$ | 5.55165 | 4.96006 | 9.21353 |
| $Z\_score\_ran\_q^2$ | 5.41451 | 4.98651 | 7.70877 |
| $\alpha\_ran\_r^2$ | 0.00000 | 0.00001 | 0.00000 |
| $\alpha\_ran\_q^2$ | 0.00000 | 0.00001 | 0.00005 |
| $r^2\_se$ | 0.2321 | 0.3130 | 0.3586 |
| $q^2\_se$ | 0.3381 | 0.4088 | 0.4699 |
| $pred\_r^2$ | -3.6132 | -2.3101 | -1.8381 |
| $pred\_r^2se$ | 1.4859 | 1.2587 | 1.1655 |

**Table 2 Statistical parameters of PLS, MLR And PCR**
The above study leads to the development of statistically significant QSAR model, which allows understanding of the molecular properties/features that play an important role in governing the variation in the activities. In addition, this QSAR study allowed investigating influence of very simple and easy-to-compute descriptors in determining biological activities, which could shed light on the key factors that may aid in design of novel potent molecules.

All the parameters and their importance, which contributed to the specific Antituberculosis inhibitory activity in the generated models are discussed here.
**1. PolarSurfaceAreaExcludingPandS:** This descriptor signifies total polar surface area excluding phosphorous and sulphur. Negative

Contibution of this descriptor to the model is -27.92%.

**2. MomInertiaX:** This descriptor signifies Moment of Inertia of the molecule. Negative Contibution of this descriptor to the model is -28.62%.

**3. SLogP:** This descriptor signifies most hydrophobichydrophilic distance. Negative Contibution of this descriptor to the model is -23.99%.

**4. SaaScount:** This descriptor signifies the total number of sulphur atom connected with one single bond along with two aromatic bonds. Negative Contibution of this descriptor to the model is -10.97%.

**5. SsOHE-index:** This is also an estate contribution descriptor which represents electrotopological state indices for number of OH group connected with three single bond. Negative Contibution of this descriptor to the model is -8.49%.

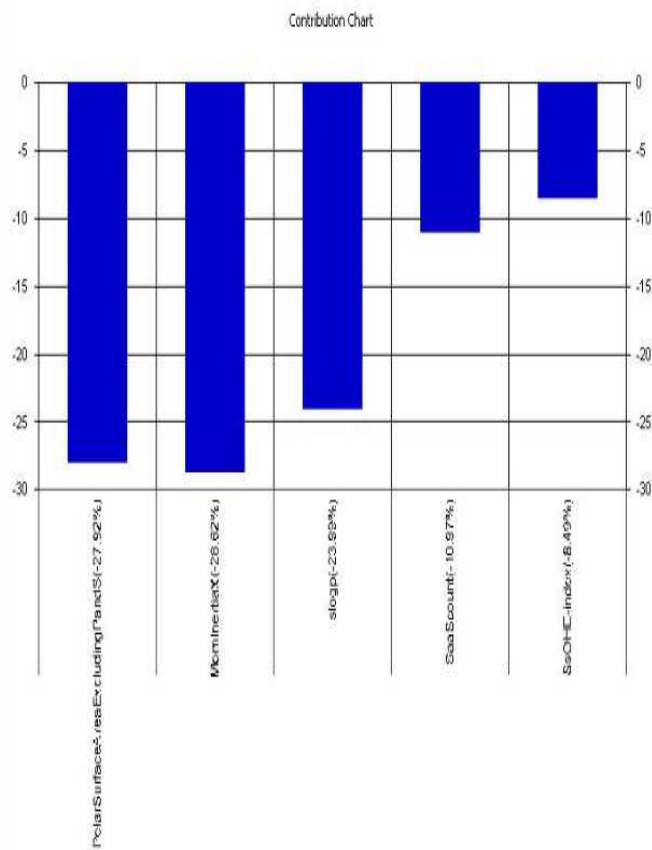

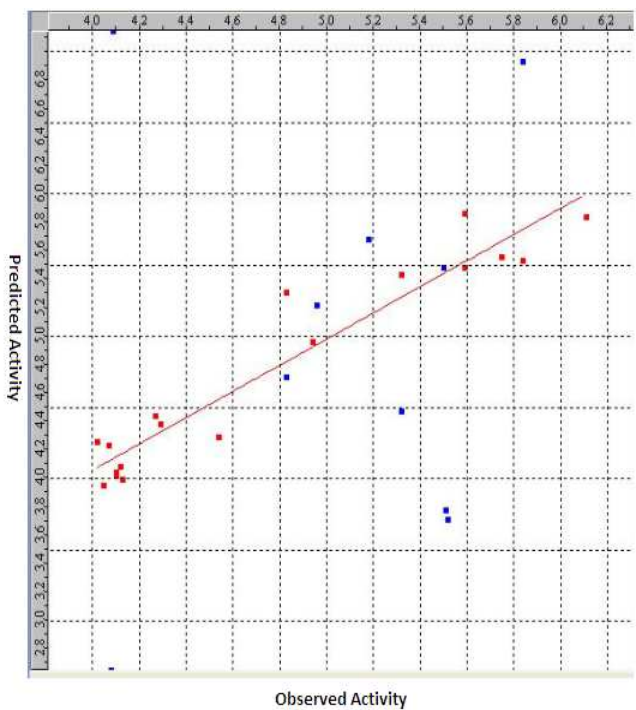

**Fig. 1** Graph of Obsered vvs. Predicted activities for training and test set molecules by Partial Least Square model. (A) Training set (Red dots) (B) Test Set (Blue dots).

**Fig. 2** Plot of percentage contribution of each descriptor in developed PLS model explaining variation in the activity

### 3.1.2 Model – 2 Multiple Linear Regression (MLR) Analysis

After 2D QSAR study by Multiple Linear Regression method using simulatead annealing variable selection method, the final QSAR equation 6 was developed having 4 variables as follows.

pIC50 = 83.7268(AveragePotential) - 0.0179(PolarSurfaceAreaExcludingPandS) - 0.00001(MomInertiaX) - 0.2923(chiV2) + 9.4202 ---(6)

Model – 2 has a correlation coefficient ($r^2$) of 0. 0.8647, significant cross validated correlation coefficient ($q^2$) of 0.7692, F test of 20.7628 and degree of freedom 13. The model is validated by $\alpha\_ran\_r^2 = 0.00001$, $\alpha\_ran\_q^2 = 0.00001$, $best\_ran\_r^2 = 0.44613$, $best\_ran\_q^2 = -0.02584$, $Z\_score\_ran\_r^2 = 4.96006$ and $Z\_score\_ran\_q^2 = 4.98651$. The randomization test suggests that the developed model have a probability of less than 1% that the model is generated by chance. The observed and predicted pIC50 along with residual values are shown in Table 1. Statistical data is shown in Table 2. The plot of observed vs. predicted activity is shown in Figure 3. The descriptors which contribute for the pharmacological action are shown in Figure 4.

All the parameters and their importance, which contributed to the specific Antituberculosis inhibitory activity in the generated models are discussed here .

**1. AveragePotential:** This descriptor signifies average of the total electrostatic potential on van der Waals surface area of the molecule. Positive Contibution of this descriptor to the model is 17%.

**2. PolarSurfaceAreaExcludingPandS:** This descriptor signifies total polar surface area excluding phosphorous and sulphur. Negative Contibution of this descriptor to the model is -24.59%.

**3. MomInertiaX:** This descriptor signifies Moment of Inertia of the molecule. Negative Contibution of this descriptor to the model is -33.57%.

**4. chiV2**: This descriptor signifies atomic valence connectivity index (order 2). Negative Contibution of this descriptor to the model is -24.83%.

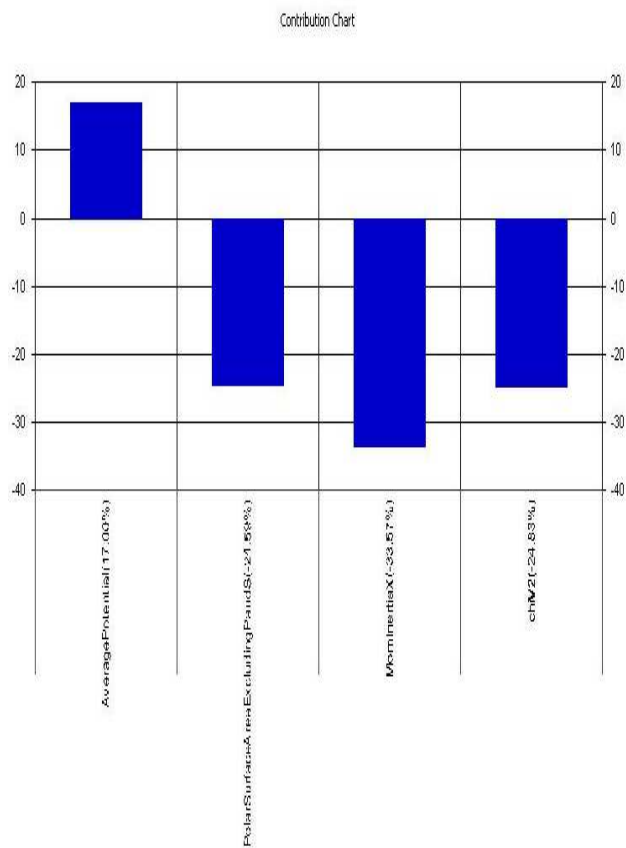

**Fig. 4** Plot of percentage contribution of each descriptor in developed MLR model explaining variation in the activity.

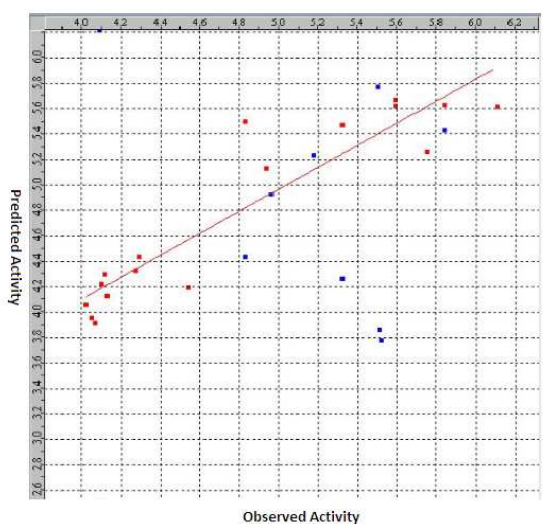

Fig .3 Graph of Observed vs. Predicted activities for training and test set molecules from the Multiple Linear Regression model. (A) Training set (Red dots) (B) Test Set (Blue dots).

### 3.1.3 Model – 3 Principal Component Regression (PCR) Analysis

The compounds were subjected to under goes PCR method to developed QSAR models by using Simulated anealining variable selection mode By using model – 3 the final QSAR equation 7 was developed having 5 variables as follows.

pIC50 = 135.3315(AveragePotential) - 0.0103(PolarSurfaceAreaIncludingPandS) - 0.0101(Quadrupole3) - 0.2208(OxygensCount) - 0.0526(HydrogensCount) + 7.6332 ---(7)

The model -3 gave correlation coefficient ($r^2$) of 0.8088, significant cross validated correlation coefficient ($q^2$) of 0.6715, F test of 19.7379 and degree of freedom 14. The model is validated by $\alpha\_ran\_r^2$ = 0.00000, $\alpha\_ran\_q^2$ = 0.00005, $best\_ran\_r^2$ = 0.25466, $best\_ran\_q^2$ = -0.13938 , $Z\_{score\_ran\_r}^2$ = 9.21353 and $Z\_{score\_ran\_q}^2$ = 7.70877. The randomization test suggests that the developed model have a probability of less than 1% that the model is generated by chance. Statistical data is shown in Table 2. The plot of observed vs. predicted activity is shown in Figure 5 .The descriptors which contribute for the pharmacological action are shown in Figure 6.

All the parameters and their importance, which contributed to the specific Antituberculosis inhibitory activity in the generated models are discussed here.

**1. AveragePotential:** This descriptor signifies average of the total electrostatic potential on van der Waals surface area of the molecule. Positive Contibution of this descriptor to the model is 27.96%

**2. PolarSurfaceAreaIncludingPandS:** This descriptor signifies total polar surface area including phosphorous and sulphur. Negative Contibution of this descriptor to the model is -19.08%.

**3. Quadrupole3:** This descriptor signifies third order magnetic dipole moments of free and bounded nucleons in the molecule. Negative Contibution of this descriptor to the model is -21.88%.

**4. OxygensCount:** This descriptor signifies total number of oxygen atoms in the Molecule. Negative Contibution of this descriptor to the model is -16.51%.

**5. HydrogensCount:** This descriptor signifies total number of hydrogen atoms in the Molecule. Negative Contibution of this descriptor to the model is -14.57%.

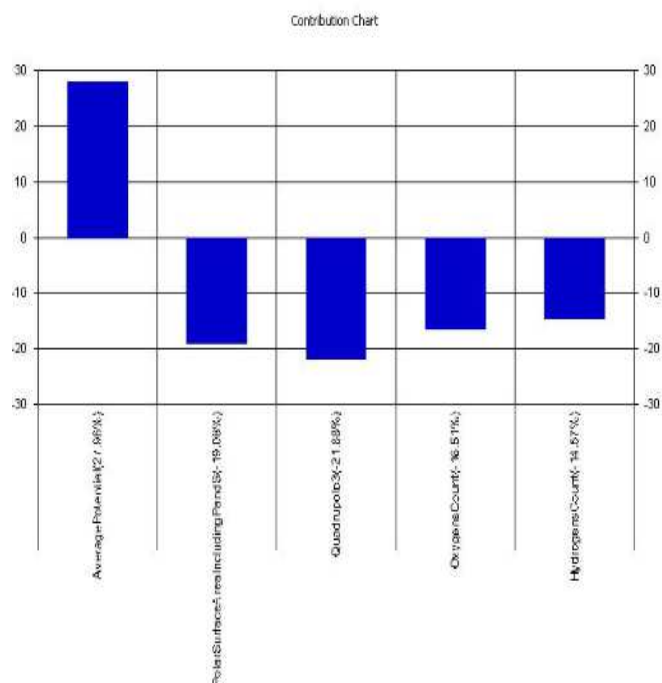

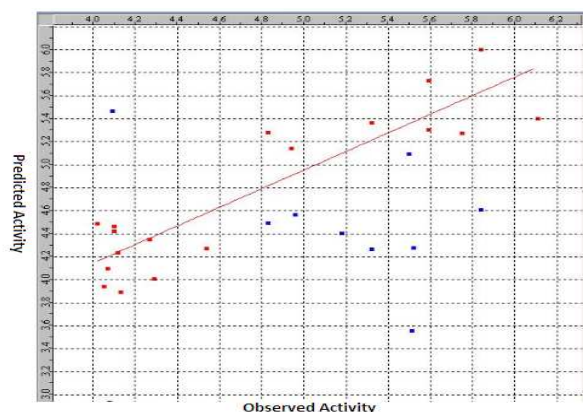

Fig. 5 Graph of Observedl vs. Predicted activities for training and test set molecules by Principal Component Regression model. A) Training set (Red dots) B) Test Set (Blue dots).

**Fig. 6** Plot of percentage contribution of each descriptor in developed PCR model explaining variation in the activity

**[IV] CONCLUSION**

The 2D QSAR studies were conducted with a series of Sulfathiazoles derivatives for Mycobacterium tuberculosis(H37Rv) inhibitors , and some useful perdictive molecular models were obtained. The physicochemical descriptors were found to have an important role in governing the change in activity. The statistical measures determine the estimation power of model for the data set from which it has been determined and evaluate it only internally. The overall degree of prediction was found to be around 86% in case of PLS,MLR and PCR. Among the three 2D-QSAR models (MLR, PCR, and PLS), results of PLS analysis showed significant predictive power and reliability as compare to other two methods.

**ACKNOWLEDGEMENTS**

The Authors are thankful to Dr Mahesh .B. Palkar Department of Pharmaceutical Chemistry K.L.E Pharmacy College Hubli.

**REFRENCES**

[1] " Tuberculosis" Centers for Disease Control and Prevention1600 Clifton Rd. Atlanta, GA 30333,

USA
http://www.cdc.gov/tb/topic/basics/default.htm
[2] "Weekly Epidemiological Record (WER)" WHO annual report on global TB control – summa ry http://www.who.int/wer/2003/wer7815/en/index.html
[3] "Update on Drug-Resistant Pathogens: Mechanisms of Resistance, Emerging Strains" by Phyllis C. Braun, PhD, and John D. Zoidis, MD http://www.rtmagazine.com/issues/articles/2004-01_01.asp
[4] "Tuberculosis management" From Wikipedia, the free encyclopedia http://en.wikipedia.org/wiki/Tuberculosis_management
[5] " Multidrug-resistant tuberculosis (MDR-TB)" From World Health Organization http://www.who.int/tb/challenges/mdr/en/
[6] "Mixed-Ligand Nickel(II) Complexes Containing Sulfathiazole and Cephalosporin Antibiotics: Synthesis, Characterization, and Antibacterial Activity" International Journal of Inorganic Chemistry Volume 2012 [2012], Article ID 106187 http://www.hindawi.com/journals/ijic/2012/106187/
[7] "Immunochemical Approaches to the Detection of Sulfathiazole in Animal Tissues" Lee N.1; Holtzapple C. K.1; Muldoon M. T.1; Deshpande S. S.2; Stanker L. H.1 Food and Agricultural Immunology, Volume 13, Number 1, 1 March [2001] , pp. 5-17(13) http://www.ingentaconnect.com/content/tandf/cfai/2001/00000013/00000001/art00001
[8] "QSAR and Drug Design" David R. Bevan Department of Biochemistry and Anaerobic Microbiology
Virginia Polytechnic Institute and State University Blacksburg, VA 24061-0308 USA http://www.netsci.org/Science/Compchem/feature12.html
[9] "Application of different chemometric tools in QSAR study of azoloadamantanes against influenza A virus" R. Karbakhsh1,* and R. Sabet2 Research in Pharmaceutical Sciences, April [2011]; 6(1): 23-33
[10] "Molecular Descriptors " The Free Online resource http://www.moleculardescriptors.eu/tutorials/what_is.htm
[11] "Molecular Descriptors Guide" Version 1.0.2 Copyright [2008] US Environmental Protection agency
[12] "Streamline Drug Discovery with CDD colabrative web based software " https://www.collaborativedrug.com/ ( Accesed in May-june [2012] )
[13] "Canv as" A comprehensive cheminformatics computing environment http://www.schrodinger.com/products/14/23/
[14] "VlifeMDS" Integrated platform for Computer Aided Drug Design (CADD) http://www.vlifesciences.com/products/VLifeMDS/Product_VLifeMDS.php
[15] "Sphere Exclusion Method for set selection" Rajarshi Guha Penn State University http://rguha.net/writing/pres/tropsha.pdf
[16] "Dissimilarity-Based Algorithms for Selecting Structurally Diverse Sets of Compounds" JOURNAL OF COMPUTATIONAL BIOLOGY Volume 6, Numbers 3/4, [1999] Mary Ann Liebert, Inc. Pp. 447–457
[17] Tropsha, A.; Gramatica, P.; Gombar,V.K. The importance of being earnest: Validation is the absolute essential for successful application and interpretation of QSPR models. QSAR Comb. Sci. [2003], 22, 69-77
[18] "On Two Novel Parameters for Validation of Predictive QSAR Models" Partha Pratim Roy, Somnath Paul, Indrani Mitra and Kunal Roy* Molecules [2009], 14, 1660-1701 ISSN 1420-3049
[19] "Multile linear Regression" http://www.ltrr.arizona.edu/~dmeko/notes_11.pdf
[20] "Variable Selection by Simulated Annealing" Dr. Frank Dieterle http://www.frank-dieterle.de/phd/2_8_6.html
[21] "Principal Components Regression With Data-Choosen Components and related methods" J.T. Gene Hwang , Dan Nettleton www.math.cornell.edu/~hwang/pcr.pdf
[22] "Partial Least Squares(PLS) Regression " Herv´e Abdi1 The University of Texas at Dallas
[23] "An introduction to partial least squares Regression" Randall D. Tobias, SAS Institute Inc., Carry, NC www.ats.ucla.edu/stat/sas/library/pls.pdf
[24] Golbraikh .A , and A. Tropsha, [2002] Predictive QSAR modeing based on diversity of sampling of experimental datasets for the training and test set selection J. Comp Aided . Mol Design , 16:357-366.
[25] "Influence of observations on the misclassification probability in quadratic discriminant analysis" https://lirias.kuleuven.be/bitstream/123456789/85608/1/qda.pdf
[26] "An introduction to the Computer Science and Chemistry of Chemical Information Systems" Craig A. James, eMolecules, Inc. http://www.emolecules.com/doc/cheminformatics-101.php

**Table 1** Structure, Experimental and Predicted Activity of Sulfathiazoles Used in Training and Test Set Using Model 1 (PLS)    Expt. = Experimental activity, Pred. = Predicted activity    a = Compound concentration in micro mole required to inhibit growth by 50%    b = -Log (IC50 X $10^{-6}$): Training data set developed using model 1 (PLS)   T = Test Set

| Sl no | Compound | IC50a(μg/ml) | PIC50b Expt | PIC50b Pred | Residual |
|---|---|---|---|---|---|
| 1 | 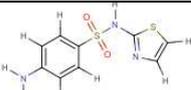 | 82.3 | 4.09 | 7.1474 | -3.0574 |
| 2 | 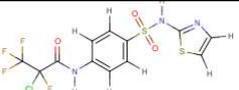 | 1.47 | 5.84 | 5.5317 | 0.3083 |
| 3 | 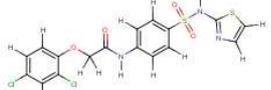 | 15.08 | 4.83 | 5.3109 | -0.4809 |
| 4 | 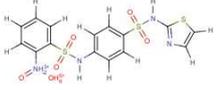 | 91.16 | 4.05 | 3.9509 | 0.0991 |
| 5 | 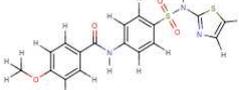 | 14.89 | 4.83 | 4.7150 | 0.115 |
| 6 | 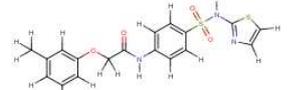 | 54.5 | 4.27 | 4.4413 | -0.1713 |
| 7 | 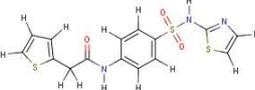 | 0.91 | 5.50 | 5.4827 | 0.0173 |
| 8 | 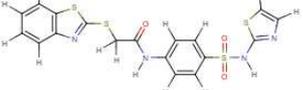 | 76.56 | 4.12 | 4.0849 | 0.0351 |
| 9 | 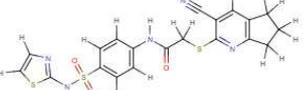 | 4.8 | 5.32 | 5.4276 | -0.1076 |
| 10 | 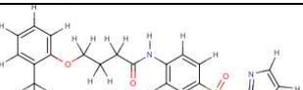 | 29.21 | 5.32 | 4.4763 | 0.8437 |
| 11 | 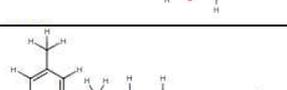 | 11.17 | 4.54 | 4.2910 | 0.249 |
| 12 | 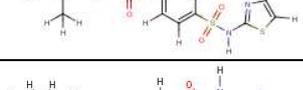 | 9.04 | 4.96 | 5.2175 | -0,2575 |
| 13 | 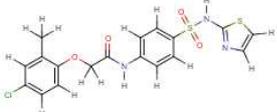 | 2.61 | 5.59 | 5.8590 | -0.269 |
| 14 | 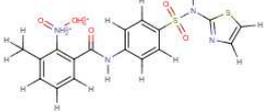 | 3.12 | 5.51 | 3.7152 | 1.7948 |
| 15 | 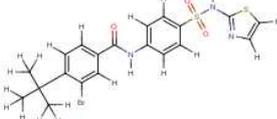 | 97.05 | 4.02 | 4.2570 | -0.237 |
| 16 | 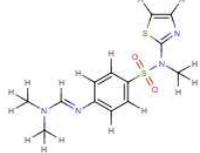 | 1.47 | 5.84 | 6.9261 | -1.0861 |
| 17 | 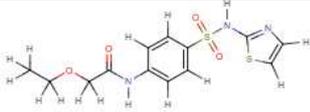 | 2.58 | 5.59 | 5.4794 | 0.1106 |
| 18 | 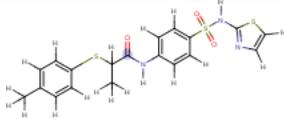 | 1.78 | 5.75 | 5.5579 | 0.1921 |
| 19 | 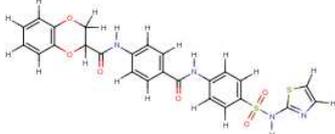 | 86.26 | 4.07 | 4.2298 | -0.1598 |
| 20 | 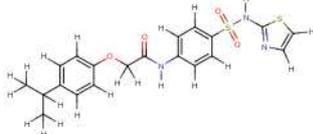 | 80.57 | 4.10 | 4.0209 | 0.0791 |
| 21 | 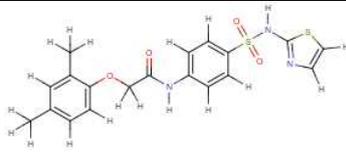 | 6.62 | 5.18 | 5.6805 | -0.5005 |
| 22 | 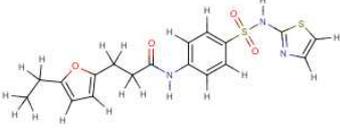 | 81.03 | 4.10 | 4.0422 | 0.0578 |

| 23 | 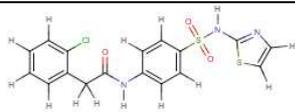 | 0.79 | 6.11 | 5.8378 | 0.2722 |
| 24 | 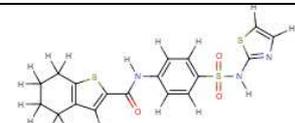 | 11.73 | 4.94 | 4.9564 | -0.0164 |
| 25 | 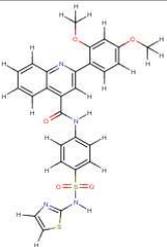 | 75.13 | 4.13 | 3.9966 | 0.1334 |
| 26 | 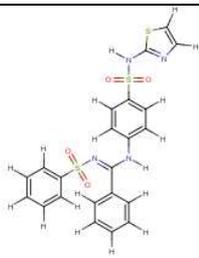 | 3.06 | 5.52 | 3.7152 | 1.8048 |
| 27 | 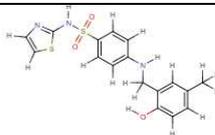 | 51.93 | 4.29 | 4.3847 | -0.0947 |
| 28 | 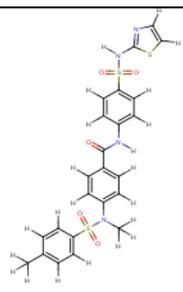 | 3.95 | 4.08 | 2.6570 | 1.423 |